\documentclass[STIX1COL]{WileyNJD-v2}

\usepackage{amsmath}
\usepackage{graphicx,color}

\articletype{Article Type}

\begin{document}

\title{A numerical approach for transient magnetohydrodynamic flows}

\author[1]{Alexander~V. Proskurin}

\author[2]{Anatoly~M. Sagalakov}

\address[1]{\orgname{Altai State Technical University}, \orgaddress{\state{Barnaul}, \country{Russian Federation}}}

\address[2]{\orgname{Altai State University}, \orgaddress{\state{Barnaul}, \country{Russian Federation}}}

\corres{*Alexander~V. Proskurin, 656038, Russian Federation, Barnaul, Lenin prospect,46. \email{k210@list.ru}}

\abstract[Abstract]{In the article the authors present a numerical method for modelling a laminar-turbulent transition in magnetohydrodynamic flows. The equations in the small magnetic Reynolds numbers approach is considered. Speed, pressure and electrical potential are decomposed to the  sum of the state values and the finite amplitude perturbations. A solver based on the Nectar++ framework is described.  The authors suggest to use of small-length local disturbances as a transition trigger. They can be imposed by blowing or electrical enforcing. The stability of the Hartmann flow and the flow in the bend are considered as examples.}


\maketitle

\section{Introduction}

Liquid metals are widely used in industry: in the blankets of advanced nuclear and thermonuclear reactors, in the large liquid metal batteries for wind and solar energy storage, in metallurgy, and so on. The facilities involved all must have a large number of pipes, valves and pumps. The magnetic field has strong influence on the liquid metal flows in them. This magnetic field can be inherent part of the technological process, as in a fusion reactor, where it holds the plasma. As is observed in nuclear reactors, the magnetic field can occur spontaneously due to the thermoelectric currents. The magnetic field can be imposed to control fluid movement as valves or pumps. For the design of all these devices, it is important to know the transition mechanisms from laminar to turbulent motion. Such studies can only be done numerically. We proposed a spectral-element magnetohydrodynamic solver  \cite{proskurin2017spectral}. The solver allows to investigate MHD flows and their stability. In this paper we propose a modification of this solver to deal with the case of nonlinear stability.

\section{Basic equations}

Consider magnetohydrodynamics equations in the form
\begin{equation}
\label{article26.NS_Syst_MagForce}
\begin{aligned}
\frac{\partial \boldsymbol{v}}{\partial{t}}+\left( \boldsymbol{v} \cdot \nabla \right)\boldsymbol{v} &= -\frac{1}{\rho}\nabla p + \nu \Delta \boldsymbol{v} + \boldsymbol{F}(\boldsymbol{v},\boldsymbol{B}_0),\\
\nabla \cdot \boldsymbol{v} &= 0,
\end{aligned}
\end{equation}
where $\boldsymbol{v}$ is the velocity, $p$ is the pressure, $\nu$ is the viscosity, $\rho$ is the density, $\boldsymbol{F}$ is the magnetic force, and $\boldsymbol{B}_0$ is the imposed magnetic field.
In most engineering applications, an induced magnetic field is very small as compared with the imposed magnetic field \cite{lee2001magnetohydrodynamic}. By neglecting the induced magnetic field, Ohm's law becomes
\begin{equation}
\label{article26.j_eq}
\boldsymbol{j} = \sigma\left( -\nabla\varphi+\boldsymbol{v}\times\boldsymbol{B}_0  \right),
\end{equation}
where $\boldsymbol{j}$ is the density of the electric current, $\varphi$ is the electric potential, and $\sigma$ is the conductivity. Using the law of conservation of electric charge ($div \boldsymbol{j} = 0$), it is possible to derive the equation for electric potential as:
\begin{equation}
\label{article26.ElPot_eq}
\Delta \varphi = \nabla \cdot (\boldsymbol{v}\times\boldsymbol{B}_0).
\end{equation}

The system (\ref{article26.NS_Syst_MagForce}) can be written in the form: 
\begin{equation}
\label{article26.WeakMHD_Syst}
\begin{aligned}
\frac{\partial \boldsymbol{v}}{\partial{t}}+\left( \boldsymbol{v} \cdot \nabla \right)\boldsymbol{v} &= -\nabla p + \frac{1}{Re} \Delta \boldsymbol{v} + St \left( -\nabla\varphi+\boldsymbol{v}\times\boldsymbol{B}_0  \right)\times\boldsymbol{B}_0 ,\\
&\nabla \cdot \boldsymbol{v} = 0,\\
&\Delta \varphi = \nabla \cdot (\boldsymbol{v}\times\boldsymbol{B}_0),
\end{aligned}
\end{equation}
where $Re = \frac{dV_0}{\nu}$ is the Reynolds number, $St=\frac{\sigma B_0^2 d}{\rho V_0}=\frac{Ha^2}{Re}$ is the magnetic interaction parameter (Stuart number), $Ha$ is the Hartmann number, and $d$, $V_0$, and $B_0$ represent the scales of length, velocity and magnetic field, respectively. This system is widely used in theoretical investigations and accurately approximates many cases of liquid metal flows 
\cite{krasnov2011comparative,lee2001magnetohydrodynamic}.

The flow variables can be decomposed to a form
\begin{equation}
\label{article26.nonlinear_form}
\begin{aligned}
\boldsymbol{v} &= \boldsymbol{U}+\boldsymbol{v},\\
\varphi &= \varphi_0+\varphi,\\
p &= p_0 +p,
\end{aligned}
\end{equation}
where $\boldsymbol{U}$, $\varphi_0$, and $p_0$ are values of a steady-state solution and $\boldsymbol{v}$, $\varphi$ and $p$ are disturbances. Now the system (\ref{article26.WeakMHD_Syst}) is:
\begin{equation}
\label{article26.disturbNS}
\begin{aligned}
\frac{\partial \boldsymbol{v}}{\partial{t}}+\left( \boldsymbol{U} \cdot \nabla \right)\boldsymbol{v}+\left( \boldsymbol{v} \cdot \nabla \right)\boldsymbol{U} + \left( \boldsymbol{v} \cdot \nabla \right)\boldsymbol{v} &=\\
-\nabla p + \frac{1}{Re} \Delta \boldsymbol{v} &+ St \left( -\nabla\varphi+\boldsymbol{v}\times\boldsymbol{B}_0  \right)\times\boldsymbol{B}_0 ,\\
\nabla \cdot \boldsymbol{v} &= 0,\\
\Delta \varphi &= \nabla\cdot(\boldsymbol{v}\times\boldsymbol{B}_0).
\end{aligned}
\end{equation}

Three types of boundary conditions may be imposed: Dirichlet, Neumann, and periodic. In the case of solid walls, zero slip boundary conditions are set for the velocity of the main flow and the perturbation
\begin{equation}
\label{article26.BounCondVWalls}
\boldsymbol{U}|_{walls} = 0, \, \boldsymbol{v}|_{walls} = 0.
\end{equation}
If walls are perfectly electrically insulated, the boundary condition is
\begin{equation}
\label{article26.BounCondPhiWallsNewmann}
\frac{\partial \varphi}{\partial \boldsymbol{n}}|_{walls} = 0
\end{equation}
both for the base flow and the disturbance. For perfectly electrically conducting walls the condition is $\varphi = const$ for the base flow and $\varphi = 0$ for the disturbance. 
At inflow and outflow the boundary conditions for the base flow and for disturbance are defined by the physical formulation of a problem. For example, for pipe and channel flows, they may be periodical.

\section{Numerical method and solver implementation}

High-order numerical methods are the most efficient tool for laminar-turbulent transition investigations. Using these methods, functions are approximated using Chebyshev, Legendre series, and so on. Such methods have good computational properties, fast convergence, small errors, and the most compact data representation. But at the present time, single domain spectral methods are applicable only for the areas of simple shapes such as rectangles, circles, triangles \cite{canuto2006spectral1}. For more complex geometry there is a spectral element method \cite{karniadakis:2013,canuto2006spectral1}. 

\begin{figure}[tb]
\begin{center}
\includegraphics[width=14cm]{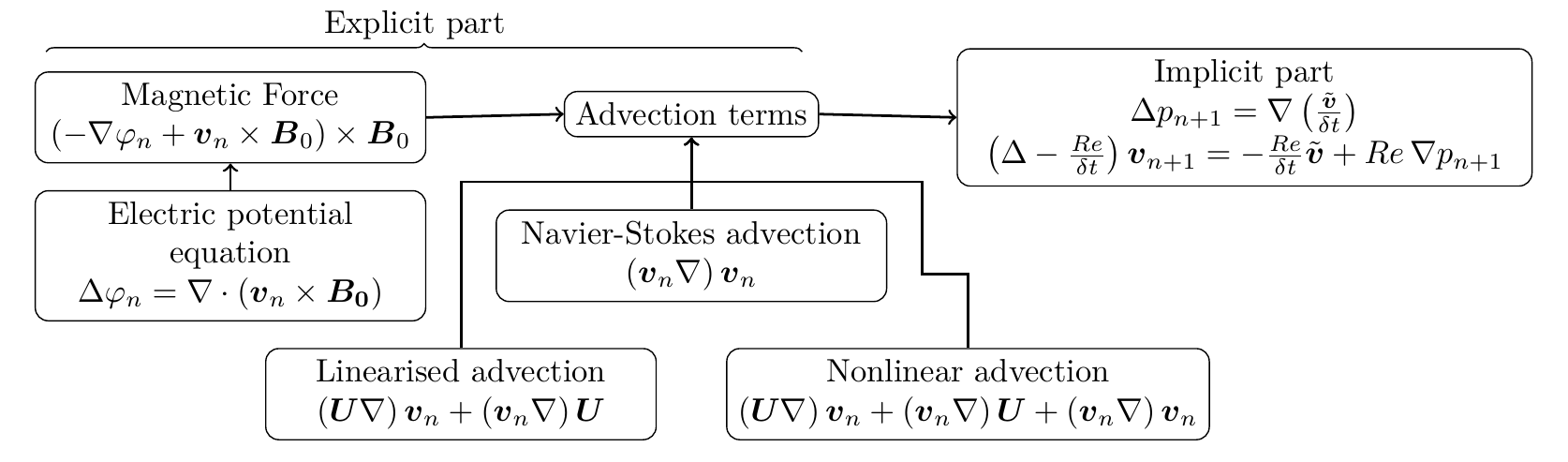}
\caption{Solver's modules diagram}\label{article26.HuntDuctMeshPict}
\end{center}
\end{figure}

Article \cite{proskurin2017spectral} introduced the MHD solver which had been developed on the basis of an open source spectral/hp element framework \texttt{Nektar++} \cite{cantwell:2015, karniadakis:2013}. In this article we describe a new feature of the MHD solver which allows modelling non-linear perturbations by equations (\ref{article26.disturbNS}). Figure \ref{article26.HuntDuctMeshPict} shows the scheme for the solver. The \texttt{MHDSolver} bases on the \texttt{IncNavierStokesSolver} from \texttt{Nektar++}, that uses the velocity correction scheme as described in \cite{karniadakis:2013,karniadakis:1991:high}. At each step it impose $\boldsymbol{v}_n$, $\boldsymbol{v}_{n+1}$ and intermediate velocity $\tilde{\boldsymbol{v}}$ in case of a first-order difference scheme for time. \texttt{Nektar++} can also  use first, second and third order implicit schemes. At the start of each step, the equation for the electric potential is solved, and the magnetic forces are calculated. The advective terms are then calculated. They can be in the usual Navier-Stokes form, in the form of linear and non-linear perturbations (\ref{article26.disturbNS}).

We investigate perturbations separately from the main flow. Such calculations are more costly than for complete flow fields, since it is necessary to calculate the advective terms three times ($\left( \boldsymbol{U} \nabla \right)\boldsymbol{v}$, $\left( \boldsymbol{v} \nabla \right)\boldsymbol{U}$ and $\left( \boldsymbol{v} \nabla \right)\boldsymbol{v}$). We made a comparative analysis and find that the additonal time is about of 20~\% of the total computation time.

However, this approach allows us to simplify data manipulation and operates narrower problem formulations for disturbances. For example, it is possible to investigate the disturbances of unstable flows. The unstable flows can be calculated using stabilizing techniques or using the steady-state solver \cite{cantwell:2015}.

\section{Example: the Hartmann flow}

\begin{figure}[hp]
\begin{center}
\includegraphics[width=10cm]{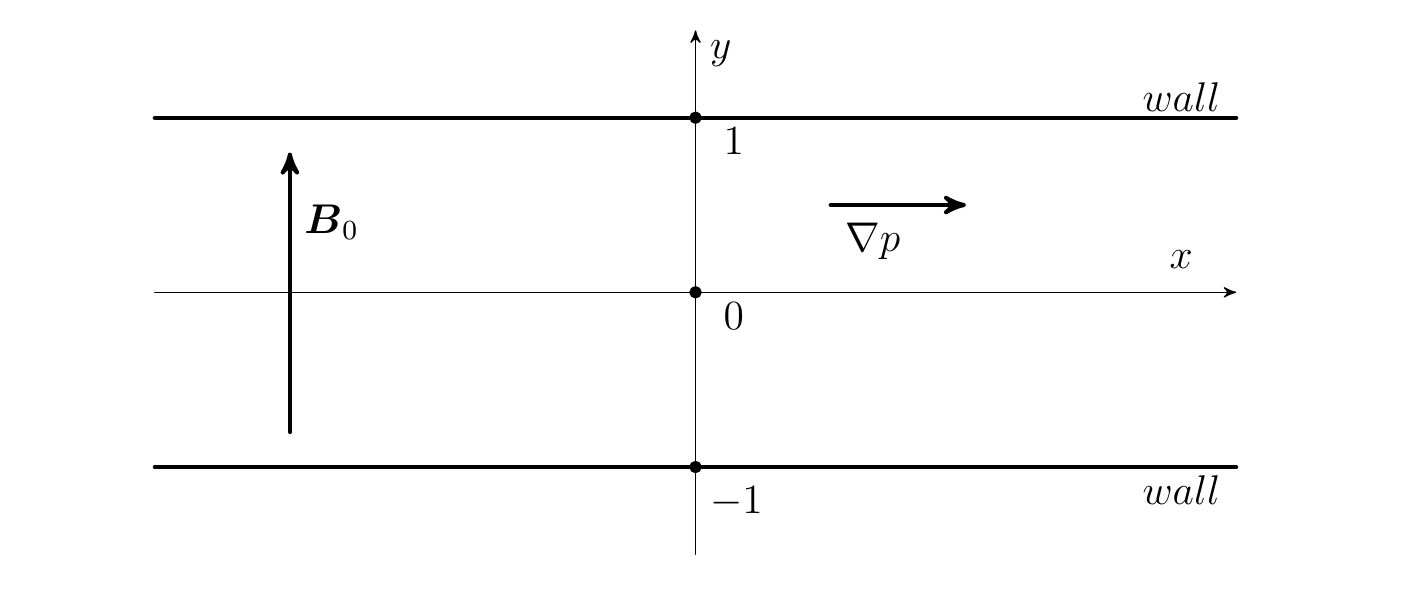}
\caption{Sketch of the Hartmann flow}\label{article26.HartmannSketch}
\end{center}

\end{figure}

As example we consider the Hartmann flow in a plane channel. The flow is shown at Figure \ref{article26.HartmannSketch}. The channel consists of two infinite parallel planes. Viscous electrically-conductive fluid flows under the action of a constant pressure gradient along the $x$ axis. The magnetic field direction is perpendicular to the planes. The base flow is
\begin{equation}
\label{article26.HartmanProfile}
\frac{u(y)}{u(0)}=\frac{\cosh{\left (M \right )} - \cosh{\left (M y \right )}}{\cosh{\left (M \right )} - 1}
\end{equation}
where $u(0)$ is the centerline velocity.

\begin{figure}[hp]
\begin{center}
\includegraphics[width=10cm]{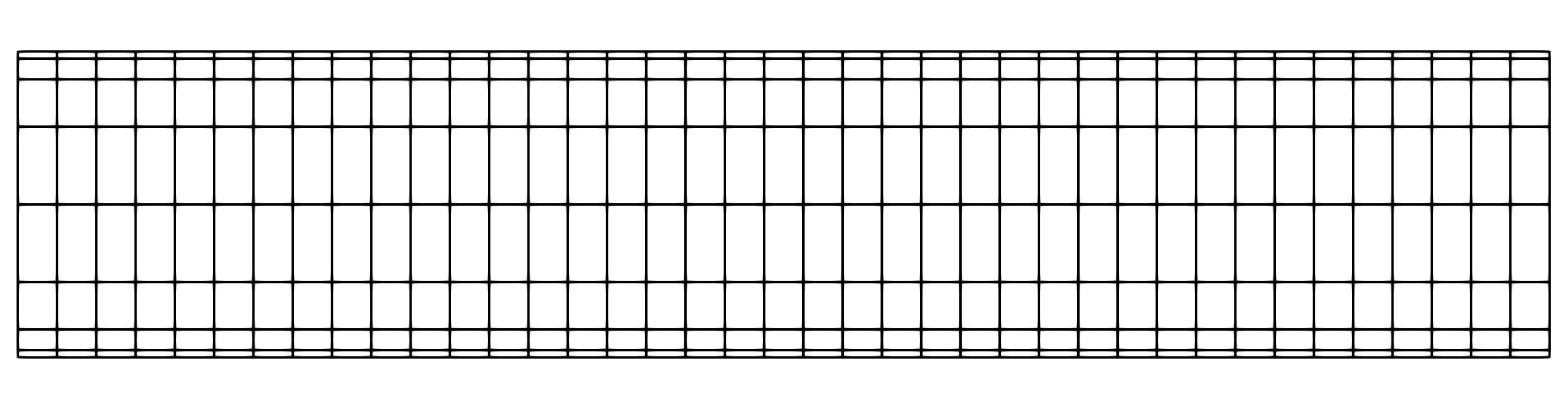}
\caption{The mesh}\label{article26.HartmannMesh}
\end{center}
\end{figure}

Figure \ref{article26.HartmannMesh} shows the mesh used for the calculations. Details of the convergence are discussed in \cite{proskurin2017spectral}. The mesh is two-dimensional and covers the $x$ and $y$ directions. The $z$ direction is supposed to be homogeneous with length $2$ and the number of Fourier series terms is $20$.

\begin{figure}[hp]
\begin{center}
\includegraphics[width=10cm]{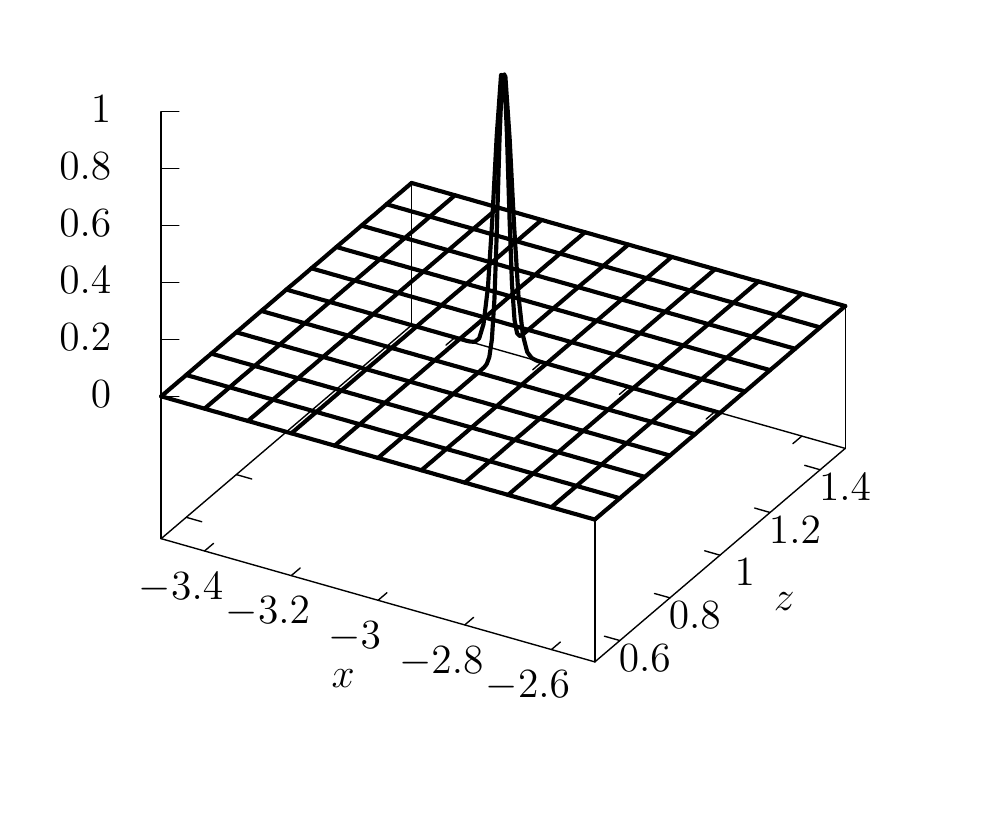}
\caption{Graph of the disturbance at the lower boundary}\label{article26.GraphInjection}
\end{center}
\end{figure}

\begin{figure}[hp]
\begin{center}
\begin{tabular}{cc}
\includegraphics[width=4cm]{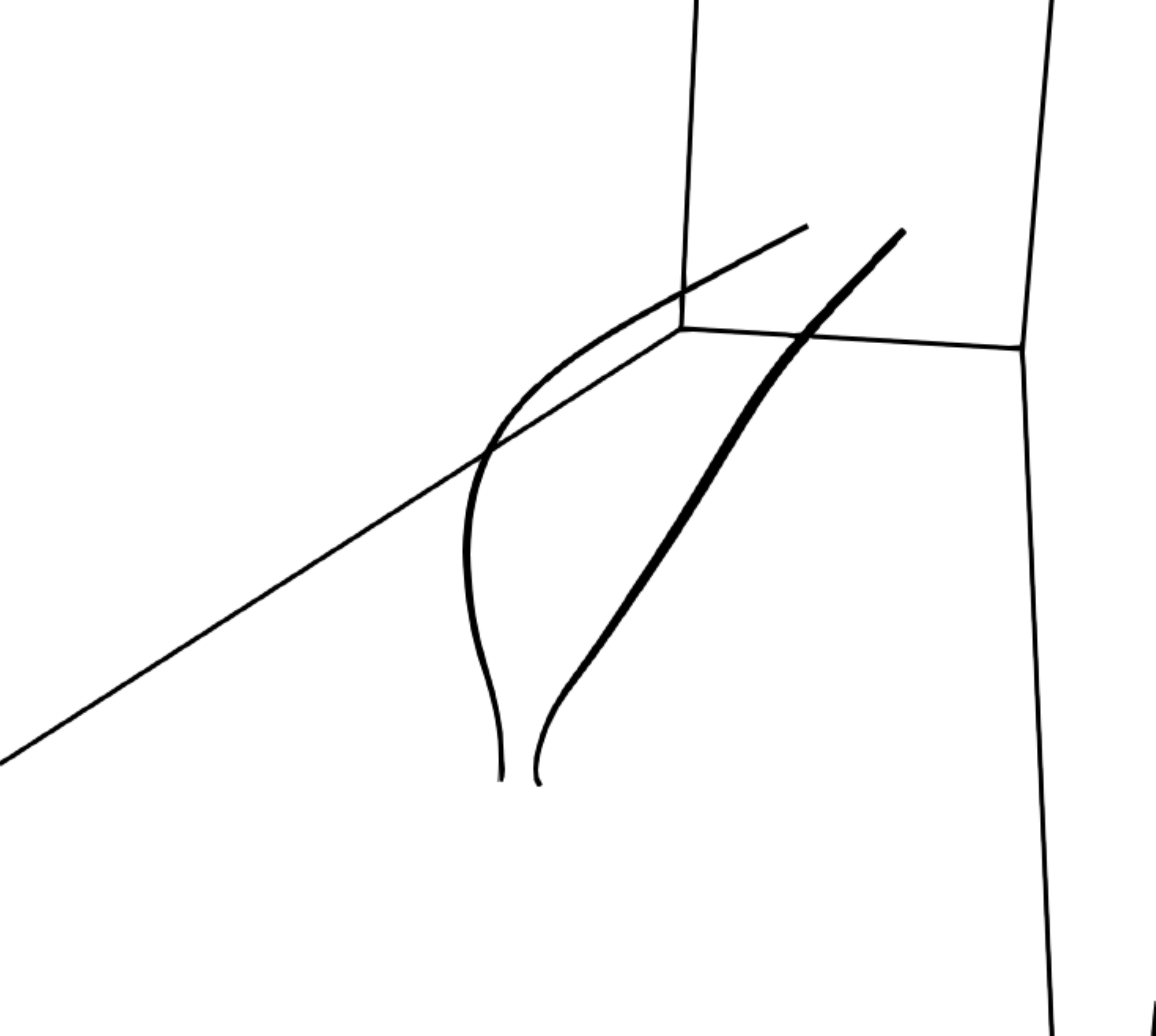}&
\includegraphics[width=4cm]{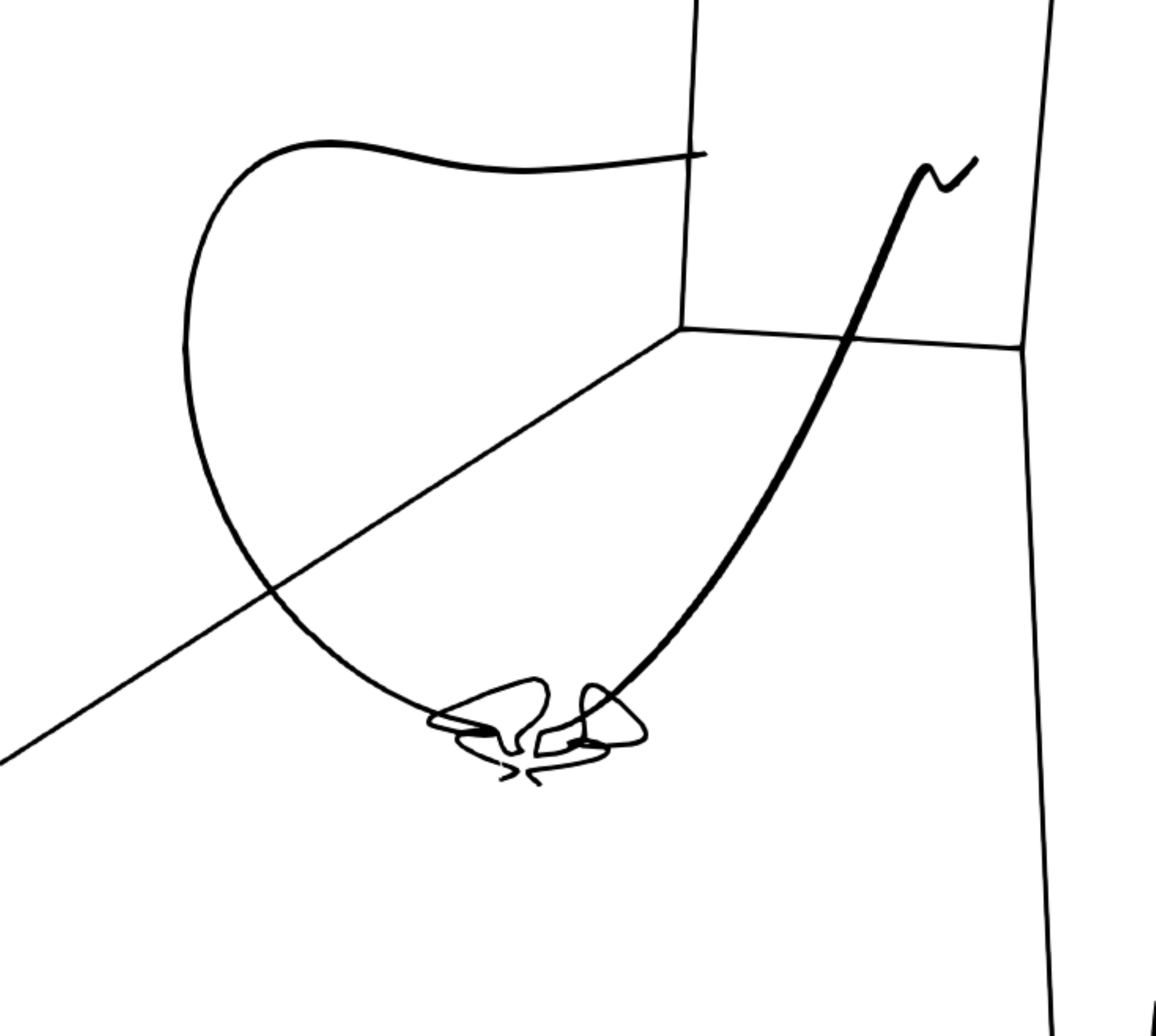}\\
(a)&(b)
\end{tabular}
\caption{Streamline samples of the disturbance at $t=0.2$ (a) and $t=1.2$ (b)}\label{article26.StreamlinesInjection}
\end{center}
\end{figure}

The simulation consists of two stages. In the first stage a disturbance is imposed on the flow. There are many ways to add disturbance to a flow. One is to use the eigenfunctions of the linearized problem \cite{orszag1980transition, proskurin2019evolution}. In article \cite{krasnov2004numerical}  an another scenario of a transition to turbulence is proposed. This scenario consists of two steps: in the first step the optimal disturbances are imposed on the base flow, and in the second step, after some time, three dimensional small-amplitude white noise is added to the disturbance. In this article, we will introduce disturbances by blowing fluid into the channel \cite{proskurin2019evolution}. To do this, on the lower plane we define the boundary condition in the form
\begin{equation}
\label{article26.Disturbance}
v_y=Ae^{(-1250.0(x + 3.0)^2)}exp^{(-1250.0*(z - 1.0)^2)},
\end{equation}
where $A$ is the maximal amplitude of the disturbance. Figure \ref{article26.GraphInjection} shows the graph of the disturbance. The energy of the disturbance depends of the amplitude $A$ and the time of the blowing $T$. The boundary condition at the upper plane and the inflow are zero, while at the outflow there is the zero friction condition. The planes are electrically insulating and boundary condition is $\frac{\partial \varphi}{\partial \boldsymbol{n}}=0$ on the upper and the lower boundaries. At the inflow and the outflow it is supposed that the disturbance of the electrical potential is zero: $\varphi=0$. Figure \ref{article26.StreamlinesInjection} shows streamlines of the disturbance at $t=0.2$ and $t=1.2$ (the total blowing time is $T=2$ and $A=1$). Symmetrical vortexes appear near the jet.

\begin{figure}[hp]
\begin{center}
\includegraphics[width=10cm]{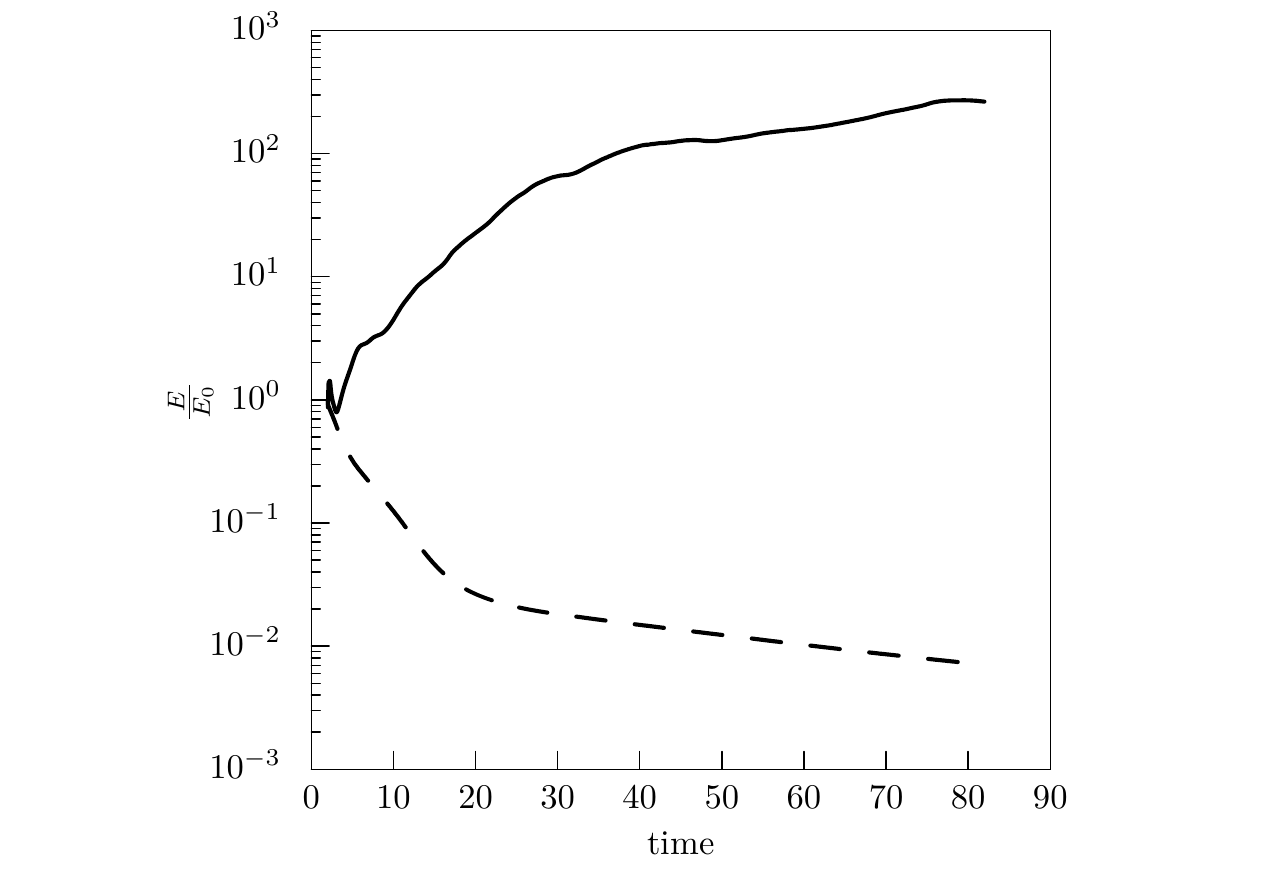}
\caption{Energy of the disturbances at $M=6$, $Re=6000$, $E_0=4.0289e-05$ (solid line); $M=15$, $Re=6600$, $E_0=7.8457e-06$ (dashed line)}
\label{article26.EnergyHartmann}
\end{center}
\end{figure}

\begin{figure}[hp]
\begin{center}
\includegraphics[width=10cm]{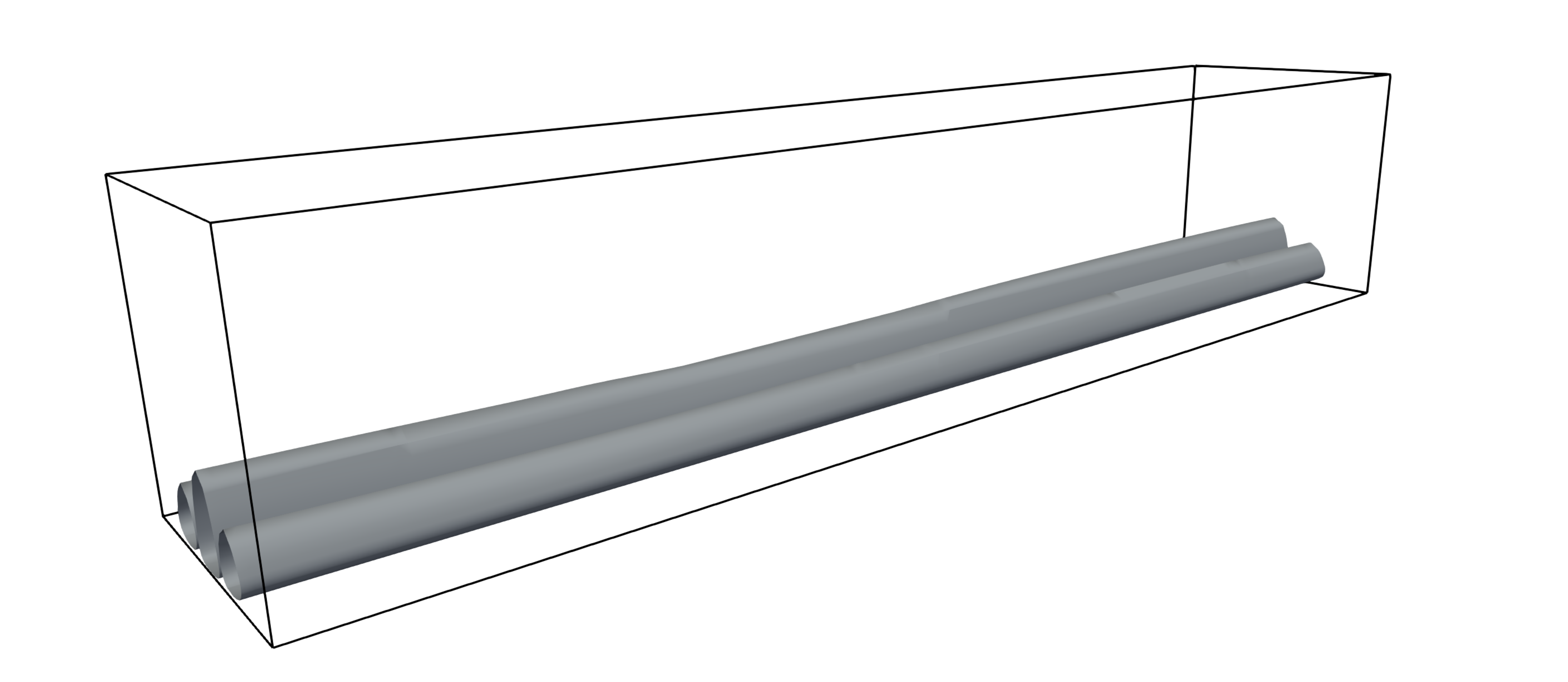}
\caption{Decreasing streaks at $M=15$, $Re=6600$, velocity amplitude isosurfaces}
\label{article26.DecreasingStreaks}
\end{center}
\end{figure}

\begin{figure}[hp]
\begin{center}
\begin{tabular}{cc}
(a) &\includegraphics[width=0.9\textwidth]{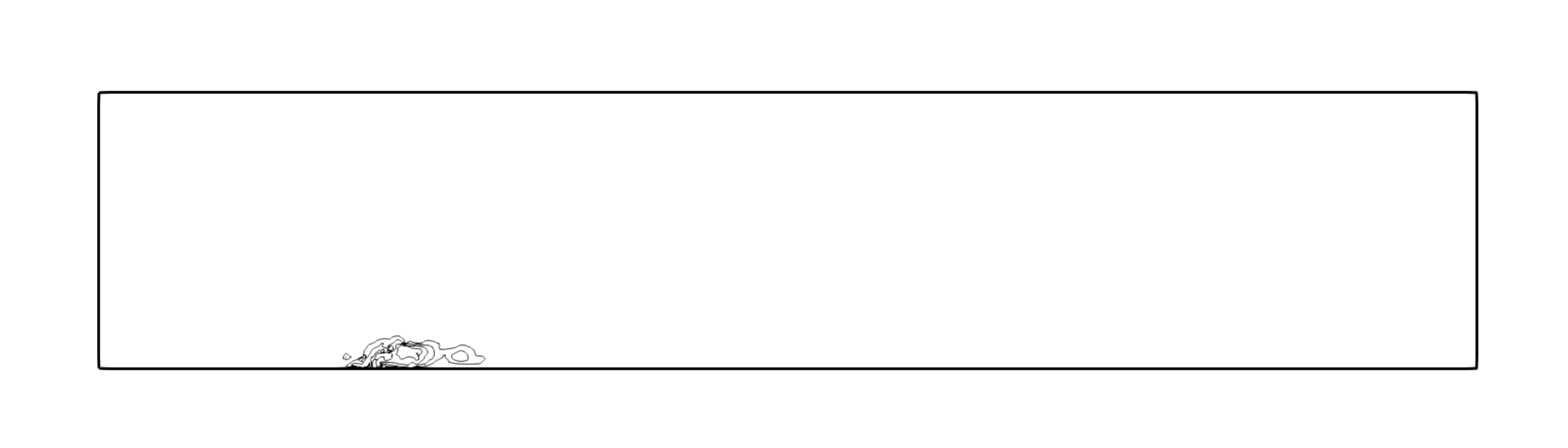}\\
(b) &\includegraphics[width=0.9\textwidth]{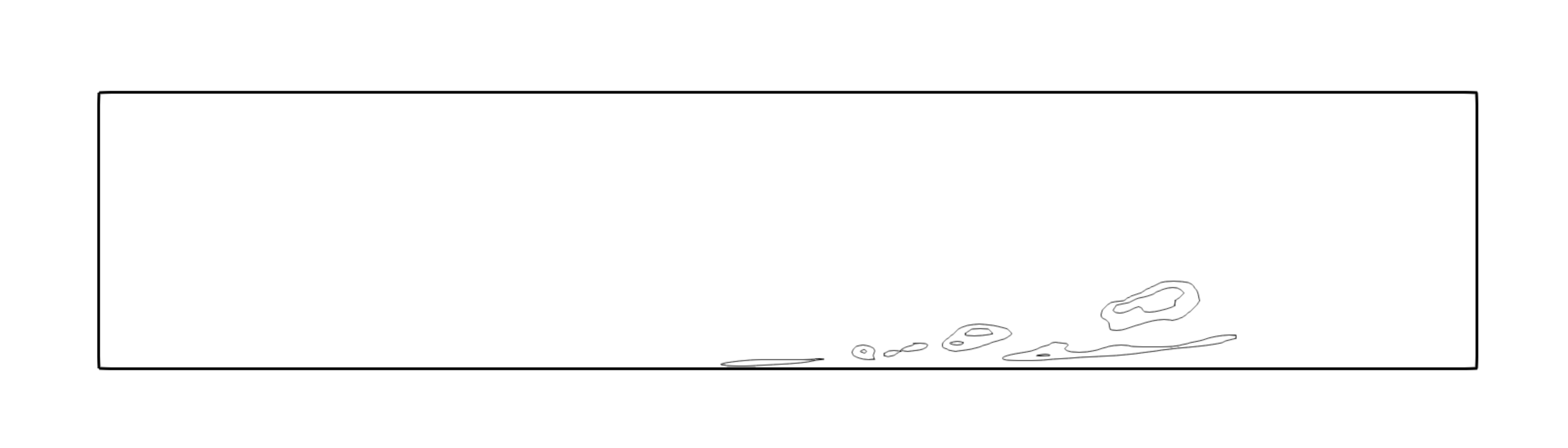}\\
(c) &\includegraphics[width=0.9\textwidth]{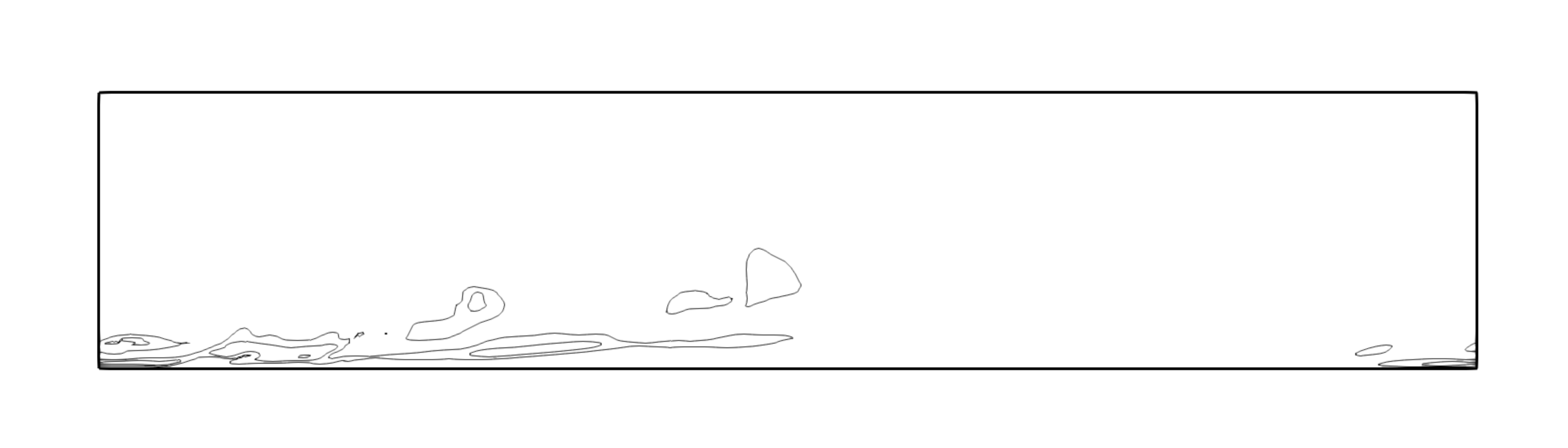}\\
(d) &\includegraphics[width=0.9\textwidth]{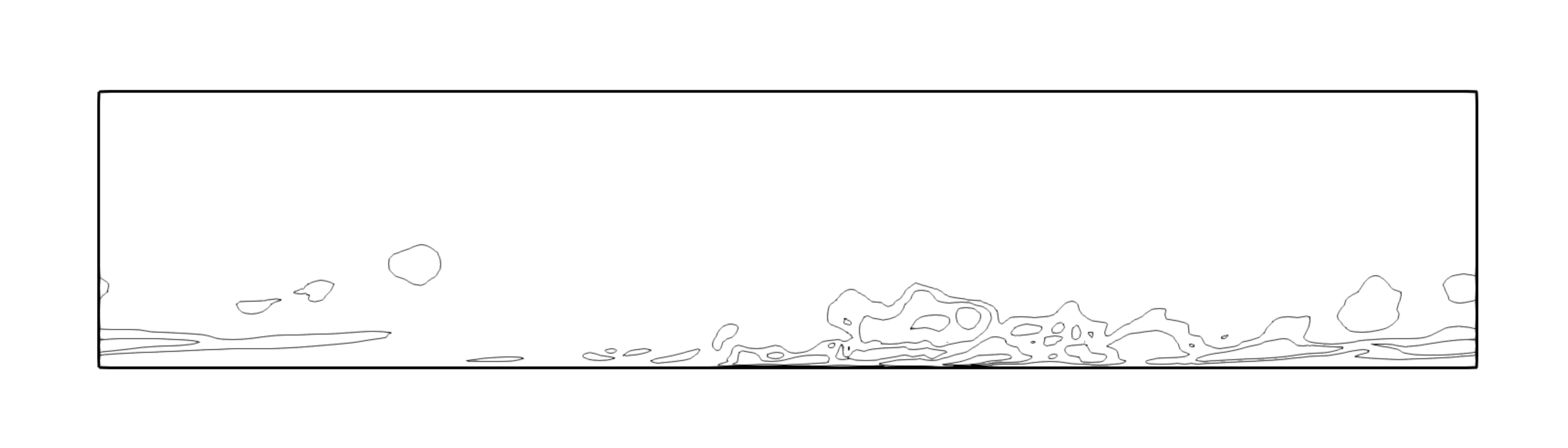}
\end{tabular}
\caption{Turbulent motion developing at $M=6$, $Re=6000$, initial condition(a) and developing in time (b),(c),(d), isosurfaces of velocity amplitude}
\label{a26.M6Re6000_isosurf_growth}
% \label{article26.HartmannGrow}
\end{center}
\end{figure}

\begin{figure}[hp]
\begin{center}
\includegraphics[width=14cm]{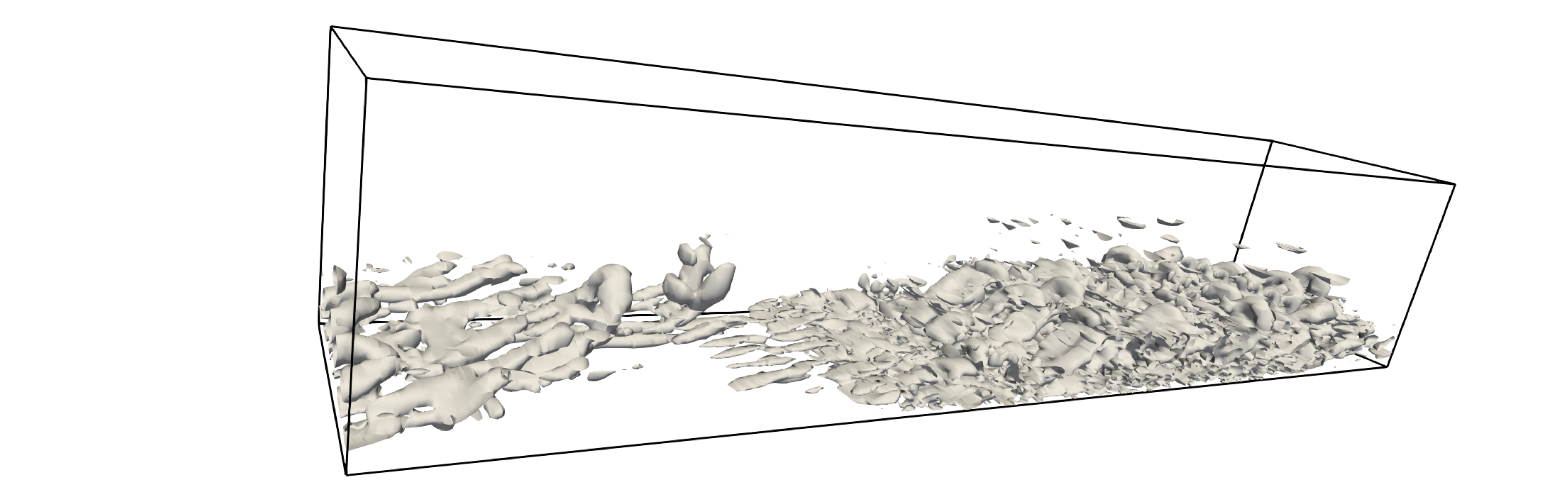}
\caption{Isosurfaces of $Q=0.1$ at $M=6$, $Re=6000$, $t=32$}
\label{article26.3d_Q0p1_disturb}
\end{center}
\end{figure}

At the second stage of the simulation the blowing is switched off and the disturbance develops under periodical conditions between the inflow and the outflow. Figure \ref{article26.EnergyHartmann} shows energy of disturbances at $M=6$, $Re=6000$, $E_0=4.0289e-05$ (solid line); $M=15$, $Re=6600$, $E_0=7.8457e-06$ (dashed line). The stability results coincide with \cite{krasnov2004numerical}. At $M=15$, $Re=6600$ the disturbance energy decreases, the flow forms the streaks (Figure \ref{article26.DecreasingStreaks}).
At $M=6$, $Re=6000$ the energy increases, the  isolines of the velocity amplitude at $t=2$(a), $12$(b), $22$(c), $32$(d) are shown in Figure \ref{a26.M6Re6000_isosurf_growth}. The plane of the figure is parallel to the axis of the channel and passes through the center of the blowing jet. Figure \ref{article26.3d_Q0p1_disturb} shows the 3D isosurfaces of the Q-criteria for Figure \ref{a26.M6Re6000_isosurf_growth}(d).

\section{Example: flow in a bend}

\begin{figure}[hp]
\begin{center}
\includegraphics[width=10cm]{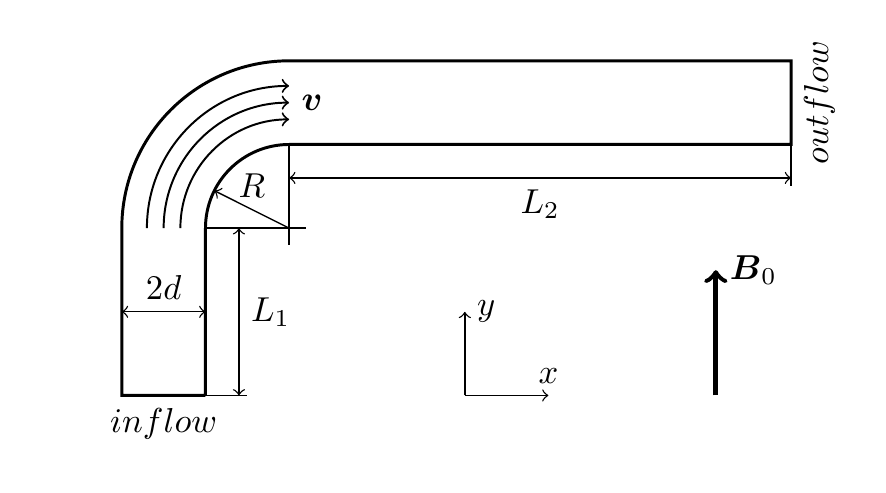}
\caption{Sketch of the bent channel}
\label{article26.L_shape_geometry}
\end{center}
\end{figure}

\begin{figure}[hp]
\begin{tabular} {cc}
\includegraphics[width=0.43\textwidth]{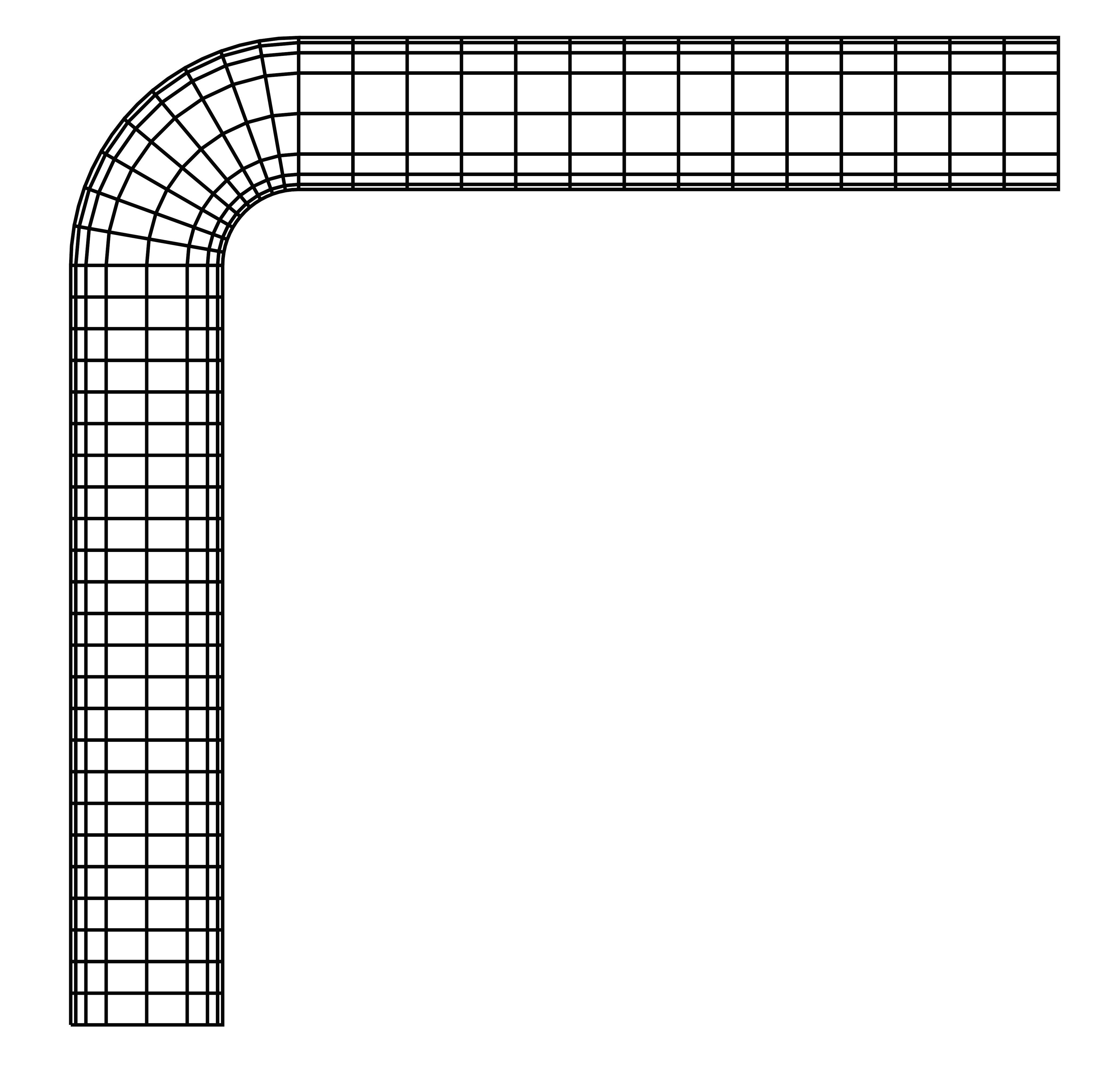}
&
\includegraphics[width=0.5\textwidth]{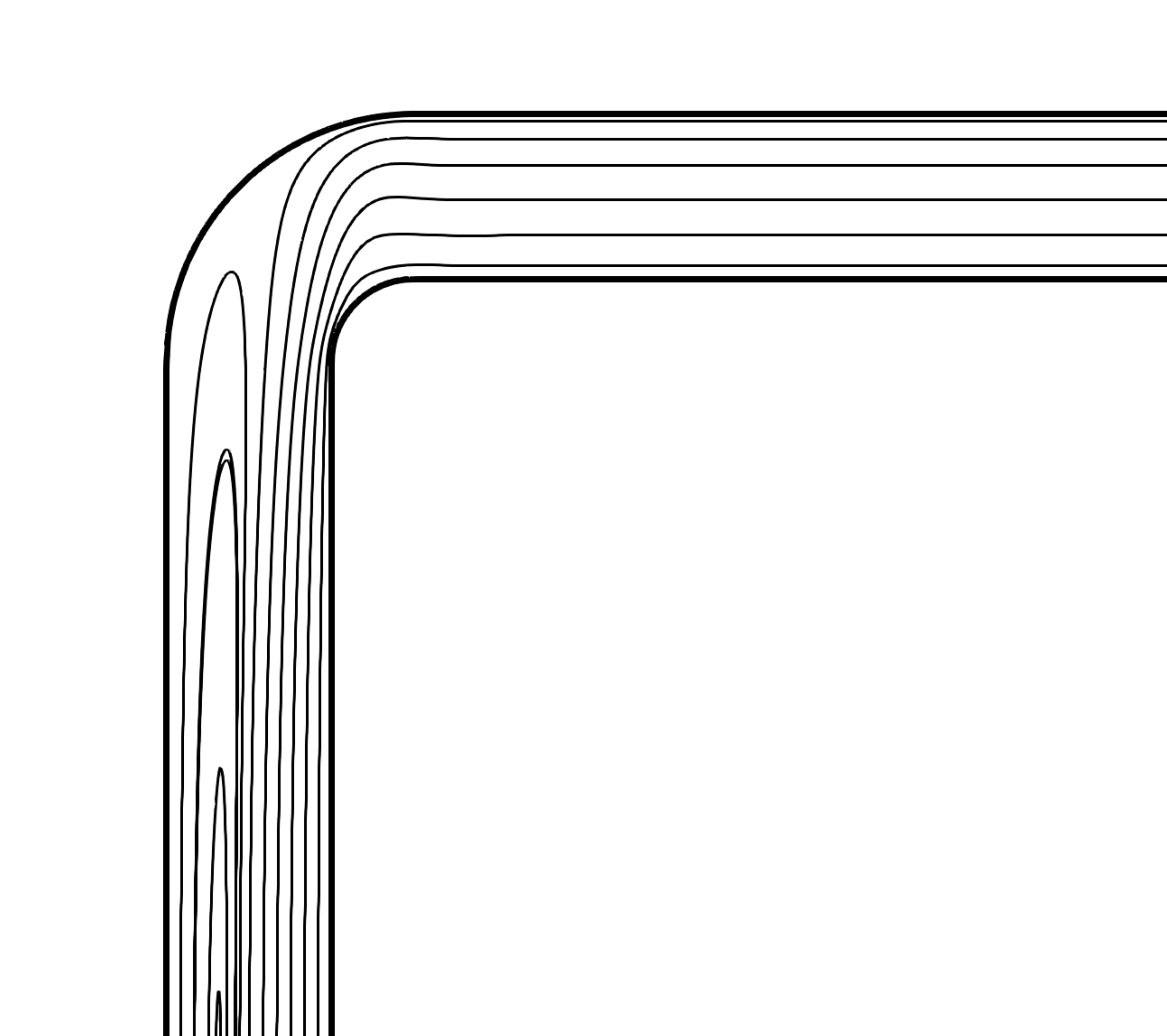}
\\
(a)
&
(b)
\\
\end{tabular}
\caption{The mesh (a) and the base flow streamlines at $M=100$, $Re=200$(b)}
\label{article26.flowM100}
\end{figure}

In a second example we consider the flow in a 90 degree bend. The state flow was investigated early in \cite{proskurin2018origin} where we found the reverse flow in the inlet branch of the channel. Figure \ref{article26.L_shape_geometry} shows the sketch of the channel: the flow is two-dimensional and the magnetic field is directed parallel to the $y$-axis. We set the Harmann velocity (\ref{article26.HartmanProfile}) as the outflow condition, and the zero friction $\frac{\partial \boldsymbol{v}}{\partial \boldsymbol{n}}=0$ as the condition at the inflow. At Figure \ref{article26.flowM100} one can see the mesh (a) and streamlines (b) of the base flow at $M=100$. 

The conditions for the disturbances differ. We set $\frac{\partial \boldsymbol{v}}{\partial \boldsymbol{n}}=\frac{\partial \varphi}{\partial \boldsymbol{n}}=0$ at the inflow and the outflow. The boundary condition on the walls is $\boldsymbol{v}=0$ and
\begin{equation}
\label{article26.DisturbancePhi}
\varphi=\pm A e^{-50.0 (y + 3.0)^2}e^{-50.0*(z - 1.0)^2},
\end{equation}
where the positive value is for the outer wall and the negative value is for the inner wall. The $z$-axis approximation is the same as for the Hartmann flow as detailed in the previous example.

\begin{figure}[hp]
\begin{tabular} {cc}
\includegraphics[width=0.5\textwidth]{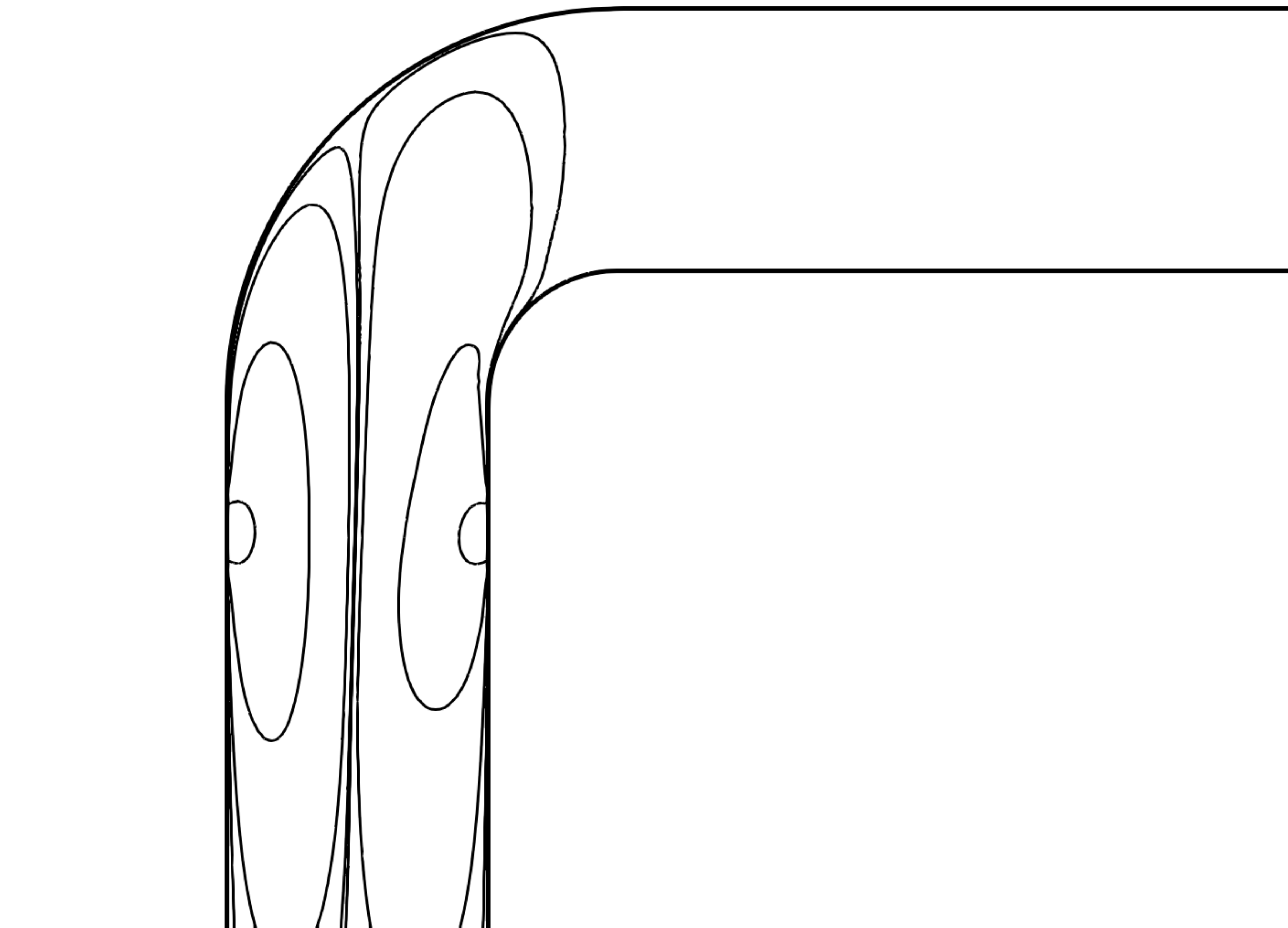}
&
\includegraphics[width=0.5\textwidth]{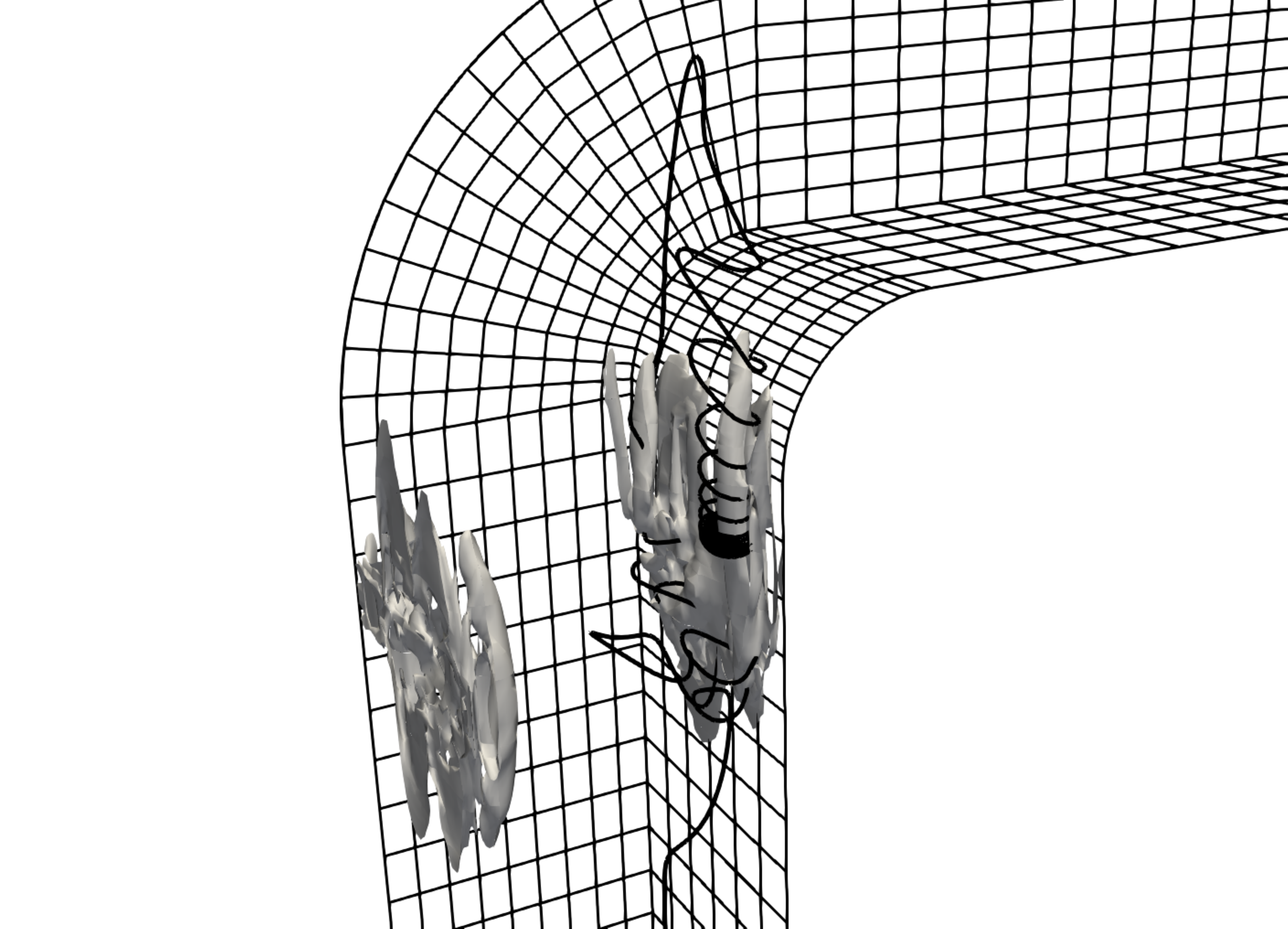}
\\
(a)
&
(b)
\\
\end{tabular}
\caption{Electric potential isosurfaces section at the center of the disturbance (a) and the disturbance vorticity isosurfaces $\omega=50$ and the streamline sample (b), all at $M=100$, $Re=200$ }
\label{article26.M100Re200initial}
\end{figure}

The isolines of the electric potential are shown in Figure \ref{article26.M100Re200initial}(a).  Physically, the expression (\ref{article26.DisturbancePhi}) can be interpreted as a small conductor separated from an electrically-conducting wall by a thin dielectric. It is possible to place conductors into a dielectric wall using an appropriate mesh, but such formulation is incompatible with the Fourier approximation along $z$-axis, and leads to a complete 3D problem, which may need the use of more computational resources.

%Also we try to enforce disturbance by $\varphi=Asin(2\pi \nu t)$ on the outer wall and $\varphi=0$ on the inner wall as done in \cite{moresco2004experimental}. We found that this disturbance do not lead to instability at $M=100$, $Re=200$ in the homogeneous numerical set-up. In the experiments \cite{moresco2004experimental} disturbances origin by the finite width electrodes. Such calculations is also 3D and more expensive than the present. 

\begin{figure}[hp]
\begin{center}
\includegraphics[width=10cm]{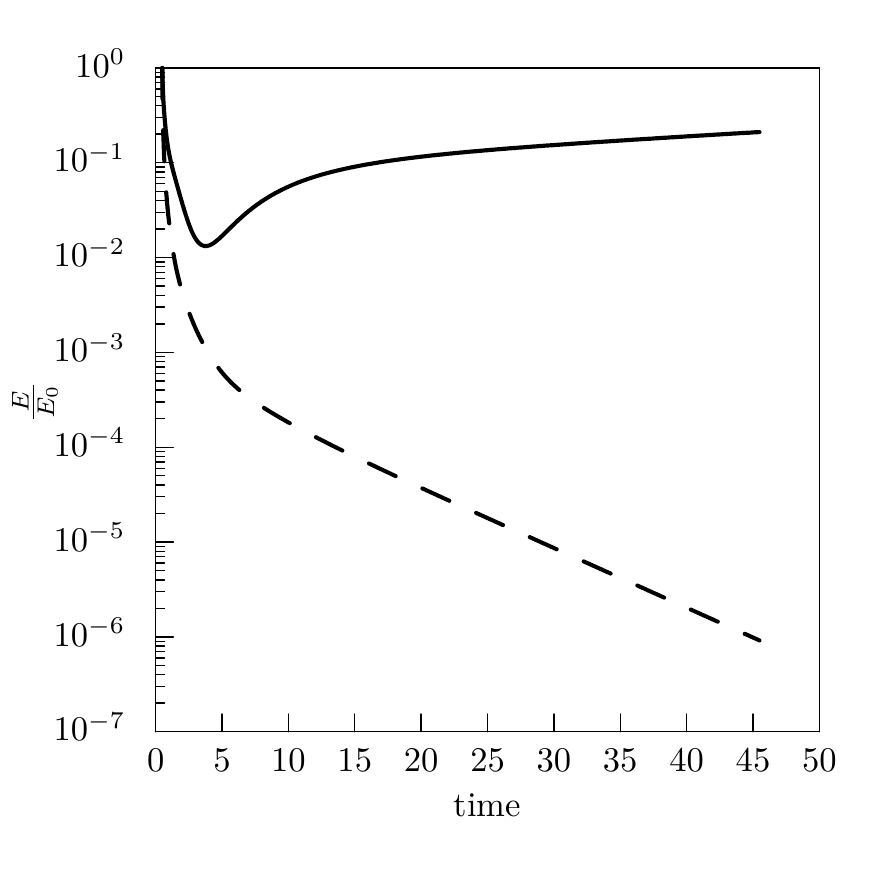}
\caption{Energy of the disturbances at $M=100$, $T=0.5$: $Re=200$, $A=8.0$, $E_0=0.19$ (solid line); $Re=100$, $A=1.0$, $E_0=0.0027$ (dashed line)}
\label{article26.EnergyBend}
\end{center}
\end{figure}

\begin{figure}[hp]
\begin{tabular} {cc}
\includegraphics[width=0.5\textwidth]{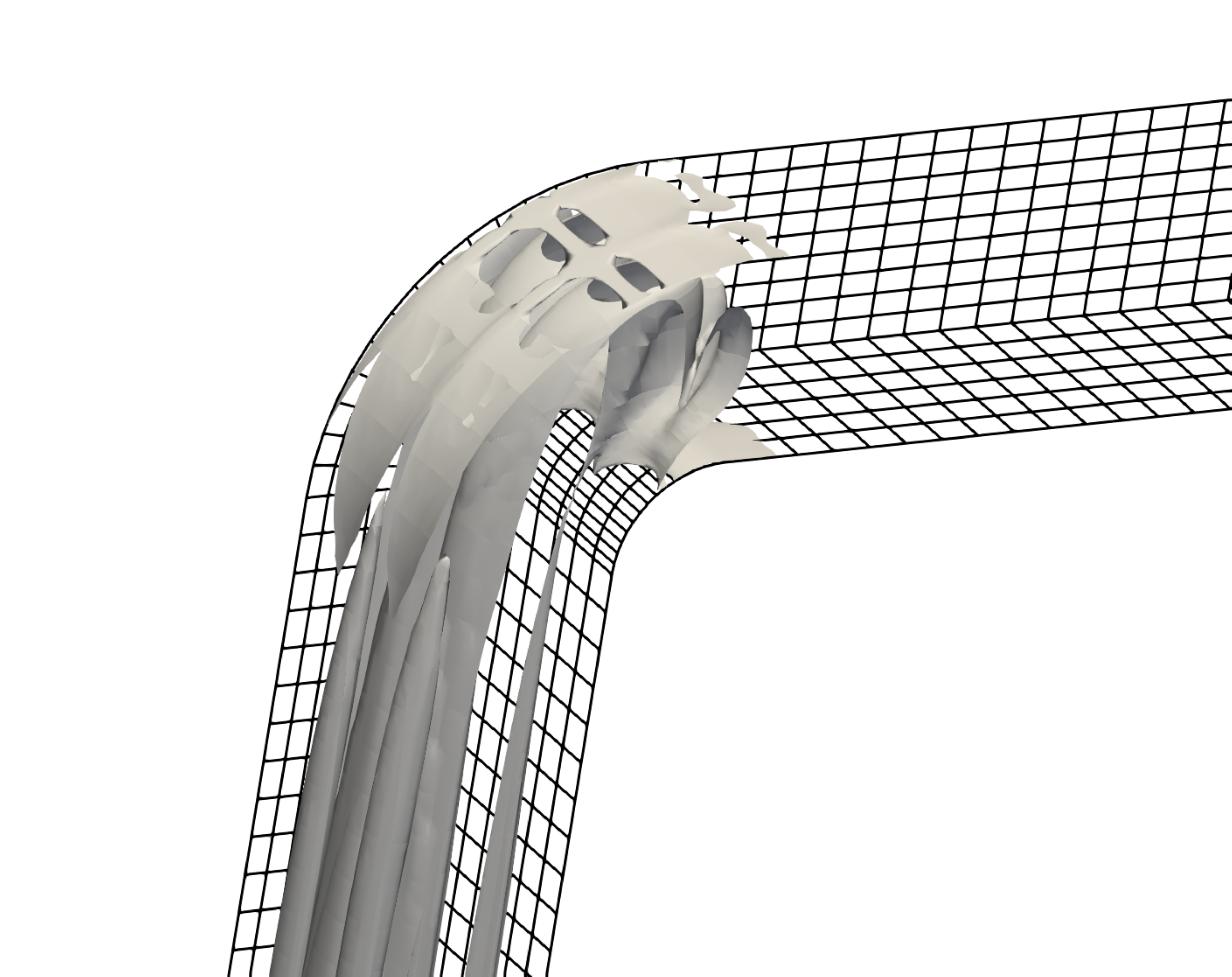}
&
\includegraphics[width=0.5\textwidth]{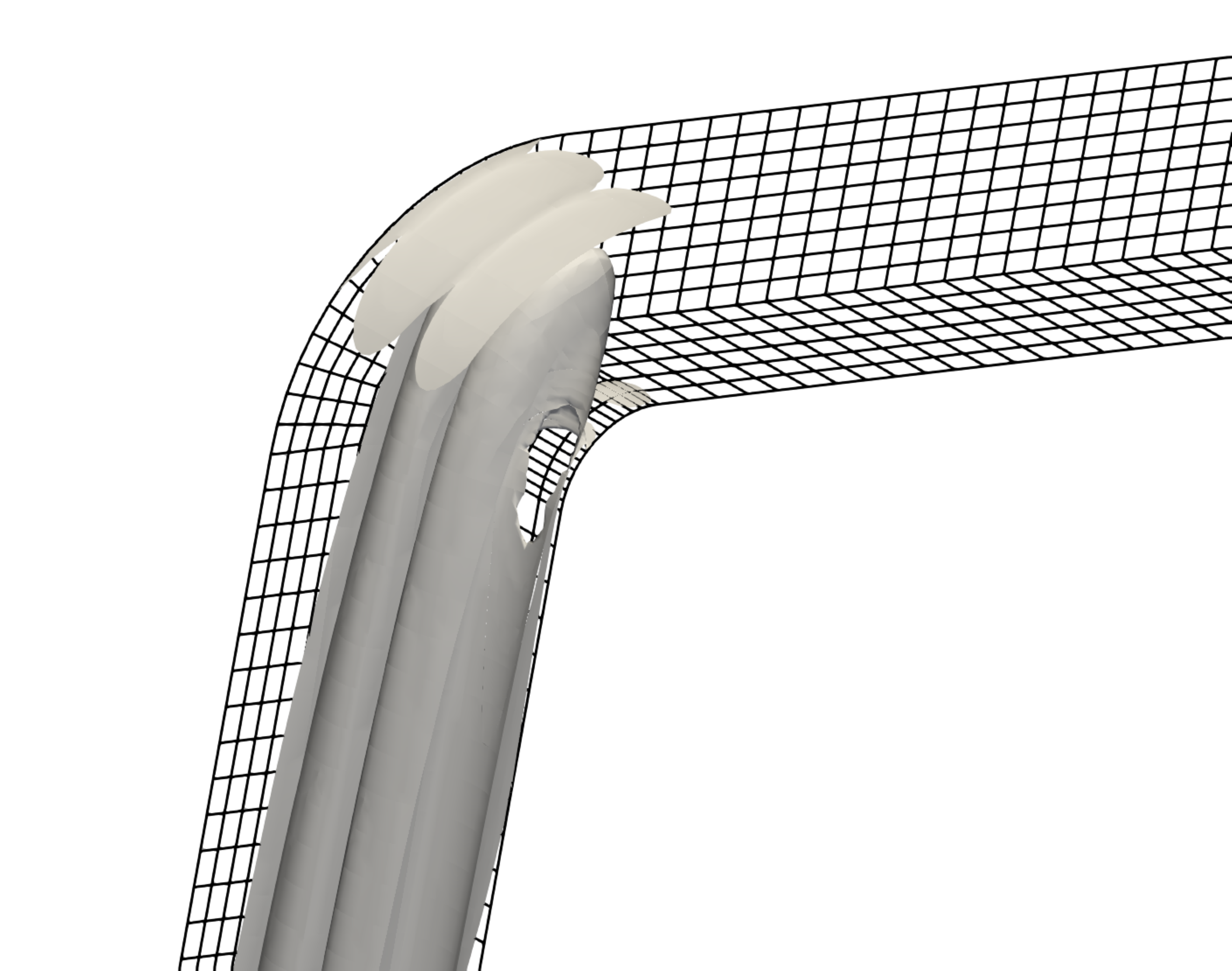}
\\
(a)
&
(b)
\\
\end{tabular}
\caption{The disturbance vorticity isosurfaces  at $t=45$, $M=100$, $Re=200$; $\omega=0.01$(a) and $\omega=0.3$(b) }
\label{article26.M100Re200grow}
\end{figure}

For the first stage we set $T=0.5$. Figure \ref{article26.M100Re200initial}(b) shows votricity $\omega=50$ isosurfaces and a streamline sample of the electrically-driven disturbance at parameter values $A=8.0$, $M=100$, $Re=200$ at the end of the first stage. Figure \ref{article26.EnergyBend} presents the second stage energy growth at $M=100$, $T=0.5$: $Re=200$, $A=8.0$, $E_0=0.19$ (solid line); $Re=100$, $A=1.0$, $E_0=0.0027$ (dashed line). At $Re=200$ we observe the energy growth but in the case of $Re=100$ the disturbance decreases. Graphs of the vorticity are shown in Figure \ref{article26.M100Re200grow} at $t=45$, $M=100$, $Re=200$; $\omega=0.01$(a) and $\omega=0.3$(b). 

%At $t=20$ disturbance patterns (a) close to eigenfunction streamlines are shown in figure (b)

\section{Conclusion}

The article proposes a novel approach to modelling the laminar-turbulent transition in magnetohydrodynamic flows. Non-linear equations for perturbations are derived. The authors present a solver that implements the approach they propose. This method is noticeably more expensive than calculations for the entire set of equations, but allows the researcher to investigate a more detailed problem set-up. For example, it is possible to set the main flow analytically, set different boundary conditions for the disturbance and the main flow, or  calculate the base flow using the steady-state solvers or the stabilization techniques. Two ways of introducing perturbations are used: blowing into the flow, and an electrical current generated disturbance. 

%Perturbations are introduced in the simplified form using the analytical expression. This form allows us to apply the Fourier expansion on the $z$-coordinate and use a desktop computer for calculations.

%The transition provocation using the linear and the optimal perturbations is quite complicated, because linear and optimal stability problems are difficult.

% Точечные возмущения являются хорошим и очень простым способом вызвать неустойчивость. Подходы, основанные на собственных функциях и оптимальных возмущениях сложнее потому что задачи линейной и оптимальной устойчивости  сложны сами по себе. 

\bibliographystyle{wileyNJD-AMA}
\bibliography{reference26}

\newpage

\end{document}